\documentclass[aps,prl,twocolumn,superscriptaddress,longbibliography,footinbib]{revtex4-2}  %
\usepackage{graphicx} % needed for figures
\usepackage{float} % locate figures, [H]
\usepackage{dcolumn} % needed for some tables
\usepackage{bm,color}
\usepackage{amssymb}
\usepackage{amsmath}
\usepackage{extarrows}
\usepackage[colorlinks=true,linkcolor=blue,urlcolor=blue,citecolor=blue]{hyperref}
\usepackage{hyperref}

\usepackage[normalem]{ulem} %makes underlining possible: \uline \uwave \sout
\definecolor{mgreen}{RGB}{1,123,0}

\definecolor{Nathanblue}{rgb}{0.,0.24,0.51}
\newcommand{\blue}{\color{Nathanblue}}

\begin{document}

\title{{\blue The cold-atom elevator: \\ From edge-state injection to the preparation of fractional Chern insulators}}

\author{Botao Wang}
\email[]{botao.wang@ulb.be}
\affiliation{CENOLI, Universit\'e Libre de Bruxelles, CP 231, Campus Plaine, B-1050 Brussels, Belgium}
\author{Monika Aidelsburger}
\affiliation{Faculty of Physics, Ludwig-Maximilians-Universit\"at M\"unchen, Schellingstr.~4, D-80799 Munich, Germany}
\affiliation{Max-Planck-Institut f\"ur Quantenoptik, 85748 Garching, Germany}
\affiliation{Munich Center for Quantum Science and Technology (MCQST), Schellingstr.~4, D-80799 Munich, Germany}
\author{Jean Dalibard}
\affiliation{Laboratoire Kastler Brossel, Coll\`ege de France, CNRS, ENS-PSL University, Sorbonne Université, 11 Place Marcelin Berthelot, 75005 Paris, France}
\author{Andr\'e Eckardt}
\email[]{eckardt@tu-berlin.de}
\affiliation{Technische Universit\"at Berlin, Institut f\"ur Theoretische Physik, Hardenbergstr.\ 36, 10623 Berlin, Germany}
\author{Nathan Goldman}
\email[]{nathan.goldman@ulb.be}
\affiliation{CENOLI, Universit\'e Libre de Bruxelles, CP 231, Campus Plaine, B-1050 Brussels, Belgium}
%\email{botao@pks.mpg.de}

\date{\today}

\begin{abstract}
Optical box traps for cold atoms offer new possibilities for quantum-gas experiments. Building on their exquisite spatial and temporal control, we propose to engineer system-reservoir configurations using box traps, in view of preparing and manipulating topological atomic states in optical lattices. First, we consider the injection of particles from the reservoir to the system:~this scenario is shown to be particularly well suited to activate energy-selective chiral edge currents, but also, to prepare fractional Chern insulating ground states. Then, we devise a practical evaporative-cooling scheme to effectively cool down atomic gases into topological ground states. Our open-system approach to optical-lattice settings provides a new path for the investigation of ultracold quantum matter, including strongly-correlated and topological phases.
\end{abstract}

\maketitle

\paragraph{Introduction.}
Optical box traps have been demonstrated as a powerful tool in cold-atom experiments~\cite{2021Navon}.
%Boxes of different shapes and dimensionalities have been realized for Bose, Fermi gases and even ultracold molecules~\cite{2013Gaunt,2015Chomaz,2017Mukherjee,2018Hueck,2021Bause}. Unlike the traditional harmonic traps, the (quasi-)homogeneous atomic gases within a box trap can lead to novel observations like the quantum Joule–Thomson effect~\cite{2014Schmidutz}, or the recurrences of coherence in a quantum many-body system\cite{2018Rauer}.
Boxes of different shapes and dimensionalities have been realized for ultracold atoms or molecules~\cite{2013Gaunt,2015Chomaz,2017Mukherjee,2018Hueck,2021Bause}, which led to novel observations including the quantum Joule–Thomson effect~\cite{2014Schmidutz} and the recurrences of coherence in a quantum many-body system~\cite{2018Rauer}.
The gas homogeneity also facilitates the probe of density-related quantities, including the quantum depletion of atomic condensate~\cite{2017Lopes}, the low-energy excitation spectrum of ultracold Fermi gases~\cite{2022Biss}, and sound speed in superfluids~\cite{2016Navon,2018Ville,2019Baird,2020Patel,2020Bohlen,2021Garratt,2021Christodoulou,2021Zhang_many}.
Besides, box traps have been used for state preparation, leading to the discovery of a novel breather in a 2D Bose gas~\cite{2019Jalm}, the deterministic preparation of a Townes soliton~\cite{2021Hassani}, and the demonstration of the transition between atomic and molecular condensates~\cite{2021Zhang}.

%An important perspective offered by box traps is to study topological state of matter~\cite{2016Goldman,2019Cooper}. 
Combined with optical lattices, box potentials allow to study a well-controlled number of atoms trapped in a few lattice sites. This exquisite control opens up new possibilities, such as measuring the growth of entanglement upon a quench~\cite{2015Islam}, or revealing fundamental properties of the Fermi-Hubbard model~\cite{2016Parsons,2017Mazurenko,2018Chiu,2019Chiu,2019Koepsell,2020Vijayan,2021Gall,2023Hirthe,2022Xu} and many-body localization~\cite{2019Lukin,2019Rispoli,2023Leonard}. 
More recently, programmable box traps enabled the generation of large homogeneous systems of more than 2000 atoms, leading to large-scale quantum simulation of out-of-equilibrium dynamics~\cite{wei2022quantum,impertro2022unsupervised,wienand2023emergence}. In the context of topological matter, a Laughlin-type fractional quantum Hall (QH) state has been recently realized in a small box filled with strongly interacting bosons~\cite{2022Leonard}. Isolating 1D lattices also allowed for the observation of the symmetry-protected Haldane
phase~\cite{2022Sompet} and 1D anyons~\cite{2023Kwan}.
Furthermore, the ability of creating optical boxes with sharp boundaries offers an ideal framework to study topological edge modes~\cite{2023Braun,2023Yao}.

\begin{figure}
	\centering\includegraphics[width=0.95\linewidth]{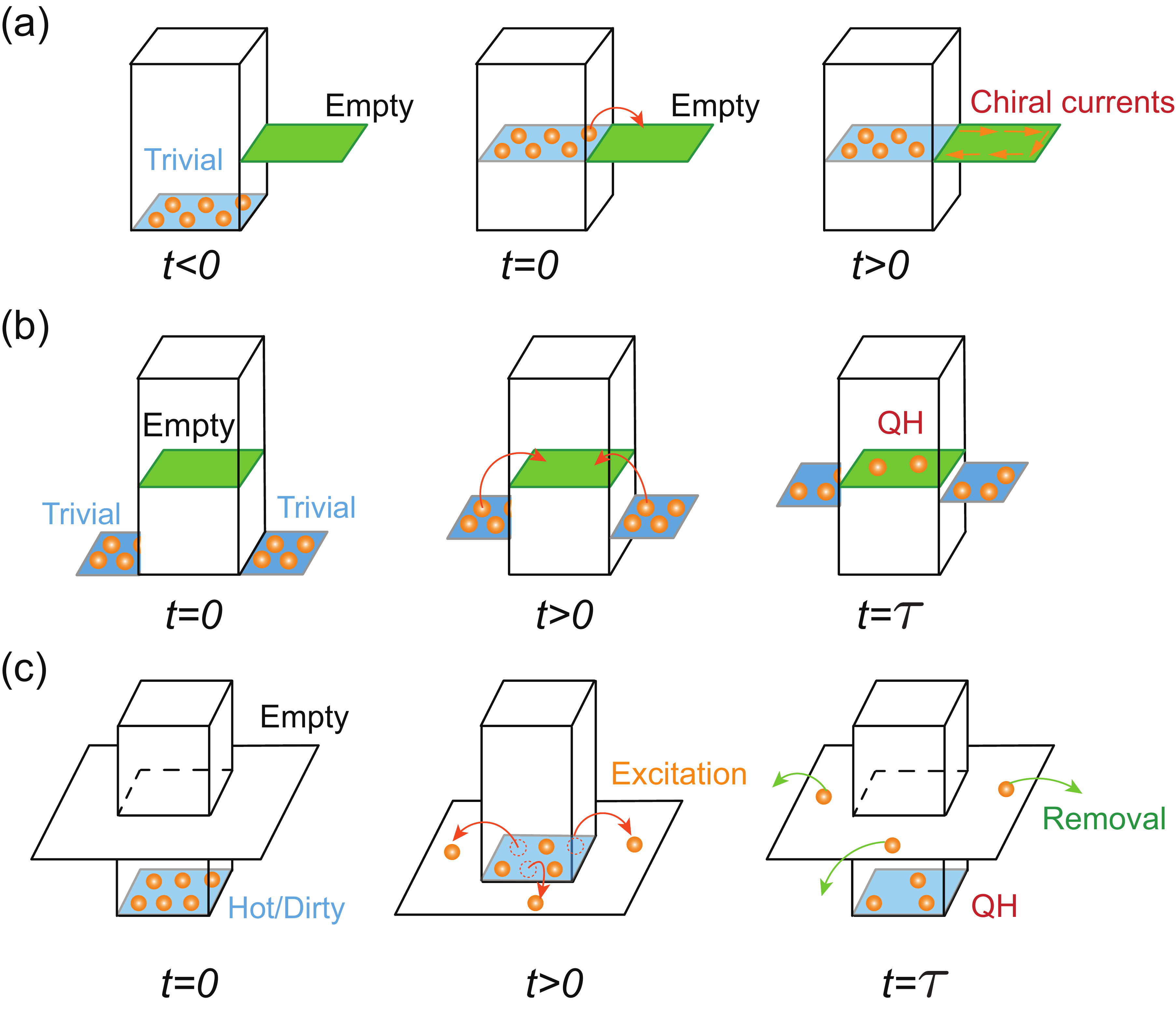} %fig_sketch_1.eps
	\caption{Sketch of the cold-atom elevator. (a) Protocol for chiral edge state injection. Setting the reservoir energy on resonance with the system's edge modes,
    particles are continuously injected into edge states in an energy-selective manner, and chiral edge currents appear in the system without populating the bulk. (b) Injection protocol for state preparation. Starting from a trivial state in the reservoirs, the latter are slowly lifted so as to adiabatically inject particles into the system until an insulating state, e.g.~a quantum Hall (QH) state, is formed. (c) Cooling protocol for state preparation. A proper tuning of the reservoir energy can be used to retrieve excitations (hot atoms) from the system. Removing the particles from the reservoir, and repeating this lift-removal process over many cycles, leads to the preparation of the desired insulating (QH) state in the system.} 
	\label{fig_sketch}
\end{figure} 

%programmable optical potentials enabled by e.g.\ digital micromirror device~\cite{2016Zupancic,2019Tajik,2021Zou},

%the logic would be based on the geometry of the sub-box configuration ---> the "reservoir" box naturally connects to the "system" box through its edge, so it is natural to use this scheme for edge-state injection (which we would present first). Then we ask: could we inject into the bulk ? Well, yes, if the system is not too large (danger of localized states), this can work ---> we prove this for a small FCI. And then only, we present the vacuum cleaner.

Inspired by the possibility of shaping box potentials of arbitrary geometries, combined with the ability to control these dynamically, we propose to use box traps to partition a lattice system into different subregions, separating a ``reservoir" region from a ``system" of interest. We explore how dynamically tuning the relative energy between these two regions allows for the controlled preparation of interesting states within the system, a scheme coined ``cold-atom elevator". 
We investigate two main scenarii: (i) injection of particles from the reservoir to the system, so as to populate edge states in an energy-resolved manner [Fig.~\ref{fig_sketch}(a)], or to prepare a strongly-interacting topological ground state in the bulk [Fig.~\ref{fig_sketch}(b)]; (ii) controlled removal of particles from an excited state (e.g. a thermal metal), performed in a repeated ``vacuum-cleaner" manner, in view of cooling the system down to a topological insulating ground-state [Fig.~\ref{fig_sketch}(c)]; see also Refs.~\cite{cooling_bernier,ho2009squeezing}. %The latter protocol is akin to evaporative cooling.

\paragraph{Edge-state injection.}
Partitioning boxes naturally provides sharp boundaries between a system and reservoirs, an ideal platform to realize and probe topological edge states. Hallmark of topologically nontrivial states, chiral edge states have been observed in photonic systems~\cite{2013Rechtsman,2013Hafezi_edge,2019Ozawa_photon} and in cold atoms using synthetic dimensions~\cite{2015Mancini_edge,2015Stuhl,2016Livi,2017An,2019Lustig,2020Chalopin,2019Ozawa}. Despite various proposals~\cite{goldman2010realistic,stanescu2010topological,liu2010quantum,goldman2012detecting,2013Goldman,
reichl2014floquet,2016Goldman_edge,2018Wang}, the realization of real-space atomic chiral edge modes has only been reported recently~\cite{2023Braun, 2023Yao}. We now show how our sub-box geometry can be used to activate topological edge currents within an empty system, in an energy-selective manner and without populating the bulk. 
%Typically, the chiral currents are generated upon a topological/Chern insulator with filled energy bands, i.e.\ with the bulk states of the underlying bands being occupied. At first glance, it is a nontrivial task to simply create chiral edge currents in a system with empty bulks. 
% it is interesting to see that our insertion protocol can be used to prepare states with higher energy, such as the edge states of CIs.

The general idea consists in coupling an empty lattice system (potentially hosting QH states) to a  reservoir, as sketched in Fig.~\ref{fig_sketch}(a). Particles are initially prepared in the reservoir, in a state that can be chosen trivial~\cite{supp}. We then perform a sudden lift of the reservoir sub-box energy $\epsilon_R$ to a \textit{proper} position, such that the energy of the particles in the reservoir becomes resonant with that of the edge mode in the system. In this way, energy-selective edge states will be populated in the initially empty system, allowing for the observation of chiral transport on a dark background.

\begin{figure}
	\centering\includegraphics[width=0.99\linewidth]{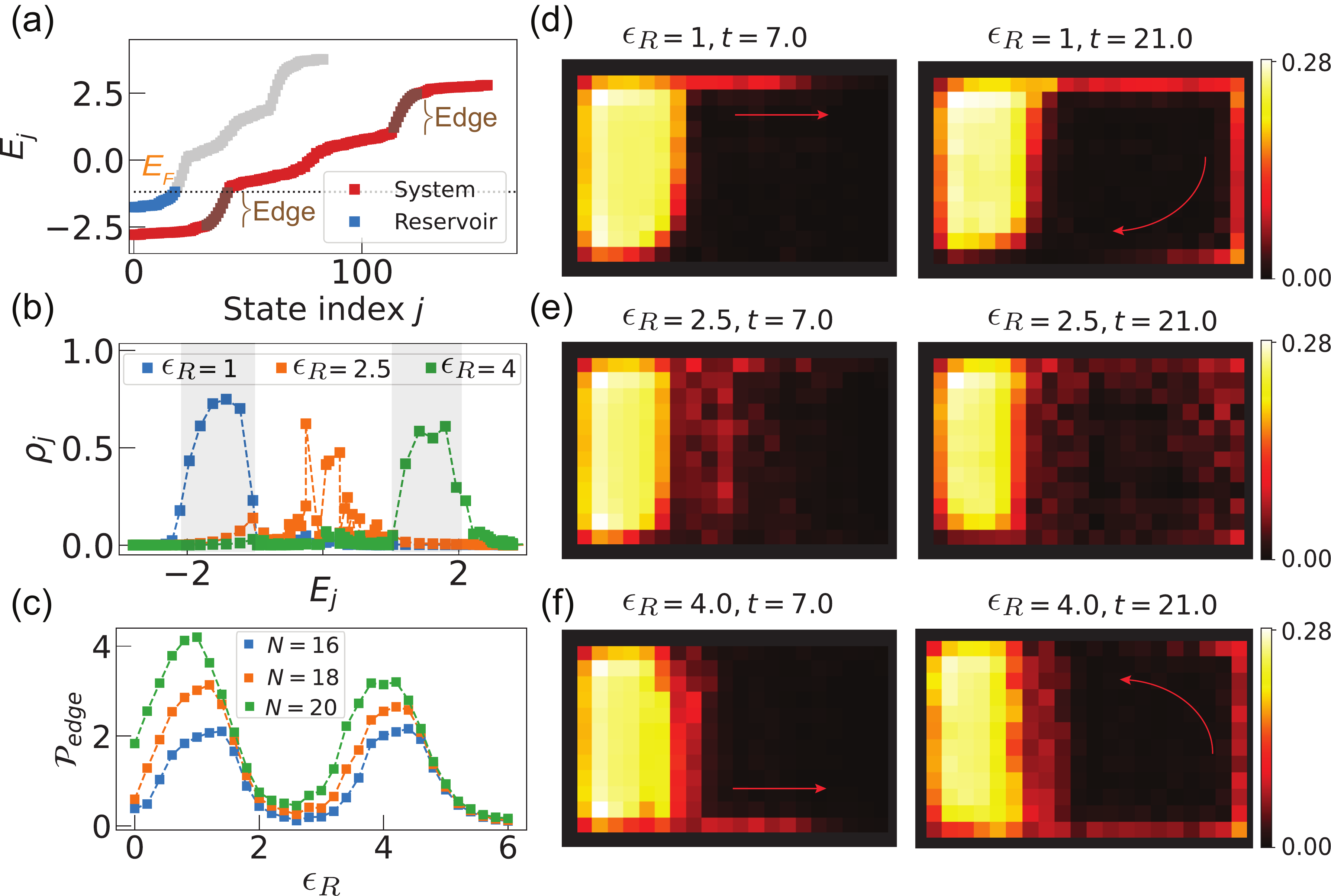} %fig_edge_1.eps
	\caption{Edge state injection in the HH model. (a) Spectrum as a function of the eigenstate index $j$. The blue and red dots correspond to the Hamiltonian describing the reservoir (with $\epsilon_R=1$) and the system, respectively. $E_F$ denotes the Fermi energy. (b) Population of HH eigenstates $\rho_j$ as a function of their energy $E_j$, at time $t=28$. (c) Edge-mode population as a function of $\epsilon_R$ for different initial particle number $N$ in the reservoir. Snapshots of spatial density distribution for (d) $\epsilon_R\!=\!1$, (e) $\epsilon_R\!=\!2.5$, (f) $\epsilon_R\!=\!4$ at times $t\!=\!7,21$. Here, a system of size $13\times 12$ and with flux $\phi\!=\!\pi/2$ per plaquette is coupled to a reservoir of size $7\times12$. Except for (c), the number of particles is $N\!=\!19$. Energy and time are in units of $J$ and $\hbar/J$, respectively. The arrows in (d) and (f) are a guide to the eye for the chiral motion.} 
	\label{fig_edge}
\end{figure}

As a concrete example, we consider the Harper-Hofstadter (HH) model~\cite{1976Hofstadter}, a square lattice with magnetic flux $\phi\!=\!2\pi \alpha$ per plaquette, coupled to reservoirs:
\begin{align}
    \hat{H}= & -{\displaystyle \sum_{\left\langle \ell\ell'\right\rangle }}\left(J_{\ell\ell'}e^{i\phi_{\ell\ell'}}\hat{a}_{\ell}^{\dagger}\hat{a}_{\ell'}+\text{h.c.}\right)+{\displaystyle \sum_{\ell}} \epsilon_\ell\hat{n}_{\ell},
    \label{H}
\end{align}%\nonumber \\ & +\frac{U}{2}{\displaystyle \sum_{\ell}}\hat{n}_{\ell}(\hat{n}_{\ell}-1)
where $\hat{a}_{\ell} (\hat{a}^{\dagger}_{\ell})$ are the annihilation (creation) operators on site $\ell$ and $\hat{n}_\ell\!=\!\hat{a}^{\dagger}_{\ell} \hat{a}_{\ell}$. We consider nearest-neighbor tunneling amplitudes $J_{\ell\ell'}$ and Peierls phases $\phi_{\ell\ell'}$, and set $\epsilon_{\ell}\!=\!\epsilon_R$ in the reservoir (zero otherwise).
Whether the reservoir is also subjected to the flux or not does not qualitatively change our findings~\cite{supp}, hence, for simplicity, we suppose that the entire system-reservoir setting is described by the HH model:~we set $J_{\ell\ell'}\!=\!1$ and choose Peierls phases $\phi_{\ell\ell'}\!=\!\phi n$ (resp.~$0$) for hopping along $x$ (resp.~$y$), where $n$ is the lattice index along $y$. 
%We set an overall potential  $\epsilon_{\ell}=\epsilon_{R}$ in the reservoir, maintaining $\epsilon_{\ell}=0$ in the system. 
%We use $J=1$ as the unit of energy throughout.

The HH Hamiltonian is a paradigmatic model of Chern insulators (CIs):~it hosts topologically nontrivial energy bands, which are characterized by nonzero Chern numbers~\cite{2019Cooper}. Setting open boundary conditions (OBC), the model hosts chiral edge modes within the bulk energy gaps~\cite{goldman2012detecting,2013Goldman,supp}. We show the energy spectrum for a system of size $13\times12$ and a reservoir of size $7\times12$ in Fig.~\ref{fig_edge}(a); the regions of low density of states (steeper slopes), correspond to chiral edge states. When setting the lift energy to the value $\epsilon_R\!=\!1$, the states populated in the reservoir become resonant with the chiral edge states located within the lowest bulk gap of the system.

Based on this observation, we show how to populate edge states in an energy-selective manner. We start with $N\!=\!19$ particles in the reservoir, which corresponds to a complete filling of its nearly-flat lowest Bloch band; the flatness allows for energy-selective population of system states. We investigate the quench dynamics obtained by solving the time-dependent Schr\"odinger equation, using different values of $\epsilon_R$. For $\epsilon_R\!=\!1$, a clockwise chiral edge current is clearly observed in Fig.~\ref{fig_edge}(d), where we plot the spatial density distribution at different times. When the box potential is lifted to $\epsilon_R\!=\!4$, i.e.\ when the reservoir is resonant with the edges states located in the upper gap, an opposite chiral motion occurs [Fig.~\ref{fig_edge}(f)]. Setting $\epsilon_R\!=\!2.5$, the populated reservoir states are resonant with the middle Bloch band of the system, in which case bulk states are populated in the system [Fig.~\ref{fig_edge}(e)].

%Starting with a fermionic ground states of the reservoir, i.e. by filling the low energy states with a few number $N_F$ of particles, with the presence of some positive energy offsets $\epsilon_R$ while keeping the door closed ($\epsilon_D=\epsilon_B=100$), we then suddenly open the door by setting $\epsilon_D=\epsilon_S=0$. In this case, particles will move from the reservoir into the system due to finite energy offset $\epsilon_R$. Interestingly, when the values of $\epsilon_R$ matches the energy of the edge states, the population of edge state can be observed. %, e.g.\ as shown in Figs.~\ref{fig_2}(a,c).
%we describe the quench dynamics by solving the time-dependent Schr\"odinger equation in terms of the Hamiltonian with vanishing barrier potentials. 

To quantify our edge-state injection scheme, we define the mean occupation $\rho_j(t)$ of an individual single-particle eigenstate $j$ in the HH system. As shown in Fig.~\ref{fig_edge}(b), we find dominant populations in the bulk gaps for $\epsilon_R=1$ (lower gap) and $\epsilon_R=4$ (upper gap). %, in which case the occupied states in the reservoir are in resonant with the edge modes of the system. 
%Therefore, lifting box potentials offers a way of populating edge states into the system in an energetically selective fashion.
Furthermore, we define the total edge-state population  $\mathcal{P}_{\text{edge}}=\sum_{j\in\text{Edge}}\rho_j$, where the state index $j$ runs over all edge modes. Figure~\ref{fig_edge}(c) shows the population $\mathcal{P}_{\text{edge}}$ as a function of $\epsilon_R$ at time $t\!=\!28$ for different particle numbers $N$. By lowering the number of fermions in the reservoir, one observes a smaller edge-mode population. In any case, the peak positions clearly indicate the energetic location of the edge modes (strong signal) and bulk modes (weak signal). As a corollary, our edge-state injection scheme can be used as a spectroscopic tool for atomic QH systems.

%\section{FCI preparation via injection}
\paragraph{FCI preparation based on particle injection.}
%A natural question concerns the possibility of using the injection scheme to form an insulating (QH) ground state within the bulk of the system. We first explored this scheme for a system of non-interacting fermions in view of forming a CI, and we found that it is only efficient for small system sizes. This limitation originates from the existence of eigenstates located deep in the bulk, which are essentially decoupled from the reservoir, and thus prevent the perfect filling of the target topological band~\cite{supp}. This issue can potentially be solved by considering a system with (strong) interactions, which suggests a possible route to prepare a fractional Chern insulator~(FCI):~a lattice analogue of a fractional QH state~\cite{2013Bergholtz,2013Parameswaran}.

A natural question concerns the possibility of using the injection scheme to form an insulating (QH) ground state within the bulk of the system. We first explored this scheme for a system of non-interacting fermions in view of forming a CI, and we present our findings in~\cite{supp}. Here, we demonstrate the applicability of this scheme to realize a fractional Chern insulator~(FCI):~a lattice analogue of a fractional QH state~\cite{2013Bergholtz,2013Parameswaran}. Several schemes have been proposed for realizing FCIs with cold atoms, based on the adiabatic variation of various system parameters~\cite{2004Popp, 2013Yao, 2014Kapit,2014Grusdt, 2015Barkeshli, 2017Repellin, 2017Motruk, 2017He, 2019Hudomal,2020Andrade}. Such a scheme was recently implemented to form an FCI state of two strongly-interacting bosons in a $4\times4$ lattice~\cite{2022Leonard}. We now show that an open-system approach, based on dynamically tuning box potentials, offers an alternative, potentially simpler and better, approach to prepare an FCI ground state with hard-core bosons.

\begin{figure}
	\centering\includegraphics[width=0.99\linewidth]{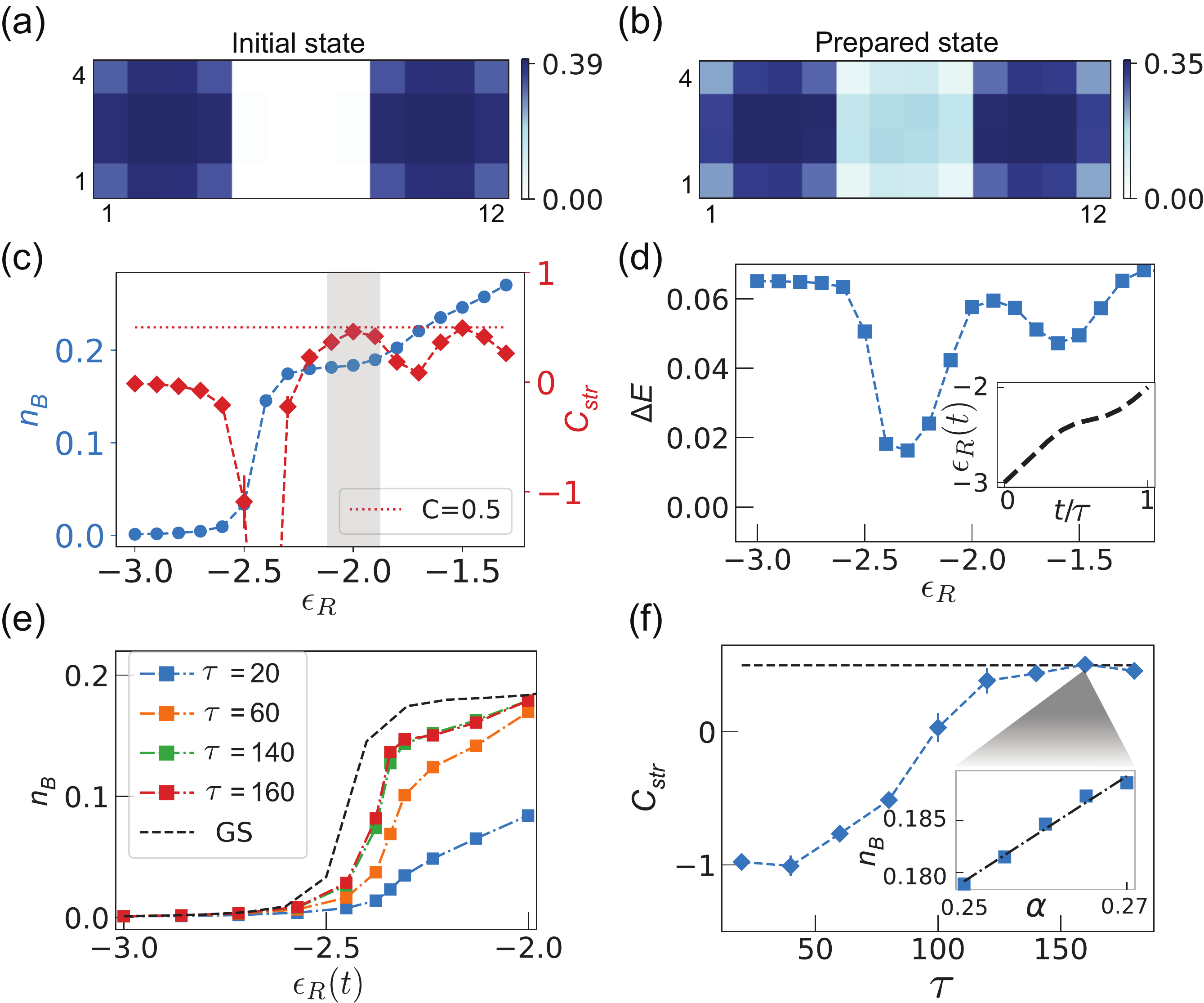}
	\caption{Preparing a fractional Chern insulator based on injection. (a) Spatial density distribution of the initial state with $\epsilon_R=-3$. (b) Density distribution of the prepared state using the ramp shown in the inset of panel (d) and $\tau=160$. (c) Bulk density and the local St\v{r}eda marker as a function of $\epsilon_R$. The shadow indicates the FCI regime. (d) Many-body energy gap as a function of $\epsilon_R$. Inset: the ramping protocol of $\epsilon_R(t)$ from $-3$ to $-2$ within time $\tau$. (e) Bulk density as a function of $\epsilon_R$ for different $\tau$ and for the instantaneous ground state. (f) Local marker as a function of $\tau$. Inset: linear fit of the density versus flux at $\tau=160$, which gives $C_{\text{str}}\!=\!0.51$. Here, we consider $N\!=\!12$ hard-core bosons; the system of size $4\times4$ is coupled to two reservoirs of size $4\times4$, with $J_R\!=\!0.15$. The error bars denote the standard error of the regression slope used to extract $C_{\text{str}}$ in Eq.~\eqref{eq_Streda}.} %C_160=0.505
	\label{fig_FCI}
\end{figure}

We consider the sub-box configuration depicted in Fig.~\ref{fig_sketch}(b):~the system is connected to two  reservoirs (without flux).
The initial state is an easily-prepared trivial state with all (interacting) particles in the reservoirs. The system region, which is initially empty, is described by the Hofstadter-Bose-Hubbard model with hard-core interactions, which is known to host a $\nu=1/2$ Laughlin-type ground state~\cite{2005Sorensen,2007Hafezi,2017Gerster,2019Rosson,2022Wang,2020Repellin}. We aim at gently injecting particles from the reservoirs to the system, by slowly lifting the reservoirs energy, in view of building up an FCI ground state in the system. 
Here, we set the hopping $J_{\ell\ell'}\!=\!J_{R}\!<\!1$ within the reservoirs and the connecting interface, to limit excitations during preparation.

We first analyze the (static) ground-state properties of our system-reservoir setup, as a function of the reservoirs' energy $\epsilon_R$. Figure~\ref{fig_FCI}(c) shows the bulk density $n_{\text{B}}$, as evaluated within the central $2\times 2$ sites. The incompressible nature of the FCI state clearly manifests as a plateau in the bulk density. In contrast to more conventional closed-system schemes, the present system automatically chooses the ideal number of bosons to form the FCI state (for the given flux value and number of lattice sites). The density reaches $n_{\text{B}}\!\approx\!0.18$ on the plateau, and we verified that it converges towards the thermodynamic prediction $n_{\text{B}}\!=\!1/8$ for increasing system sizes~\cite{supp}.
%We observe that $N_{\text{Bulk}}$ forms a plateau within the range $-2.2\lesssim \epsilon_{R}\lesssim -1.9$, which represents the incompressible bulk properties. 
As another hallmark signature of the FCI, we evaluate the fractionally-quantized Hall conductivity $\sigma_{\text{H}}$, which is encoded in the density distribution via St\v{r}eda's formula~\cite{1982Widom,1982Streda,1983Streda,2008Umucalilar,2020Repellin}, 
\begin{equation}
    C_{\text{str}}=\frac{\partial n_{\text{B}}}{\partial\alpha}=\frac{\sigma_{\text{H}}}{\sigma_{0}},\label{eq_Streda}
\end{equation}
where $\sigma_0\!=\!1/2\pi$ is the conductivity quantum. For a $\nu\!=\!1/2$ Laughlin state, the St\v{r}eda marker is expected to take the value $C_{\text{str}}\!=\!1/2$, which is the many-body Chern number of the state.
%To further characterize their topological properties, we use a local marker according to the St\v{r}eda's formula, $C_{\text{str}}=\partial n/\partial\alpha$ with $n$ being the particle density. 
In our case, we find $C_{\text{str}}\!\simeq\!0.46$ at $\epsilon_R\!=\!-2$, hence indicating the precursor of a fractional Hall response [Fig.~\ref{fig_FCI}(c)]. It is interesting to compare this result to the value $C'_{\text{str}}=0.61$, which is obtained in the experimental closed-box configuration of Ref.~\cite{2022Leonard}, where an exact number of bosons ($N=2$) is loaded in $4\times4$ sites; this comparison supports the idea that the system optimizes the formation of an FCI state when coupled to reservoirs.

The bulk density and the St\v{r}eda marker both show an interesting behavior across the transition that occurs within the system, as particles enter the system and eventually form the FCI state. Indeed, in the vicinity of $\epsilon_R\!\approx\!-2.5$, one notices an abrupt increase in $n_{\text{B}}$ and a breakdown of St\v{r}eda's formula [Fig.~\ref{fig_FCI}(c)], accompanied with a sudden drop in the many-body gap [Fig.~\ref{fig_FCI}(d)]. The minimal many-body gap associated with this transition is $\Delta\!=\!0.016$, which suggests a realistic ramping time $\tau\!\sim\!100$ for adiabatic state preparation, compatible with recent experiments~\cite{2022Leonard}.

%Monitoring the bulk density also provides a signature of (a precursor of) a topological phase transition. In the vicinity of $\epsilon_R\!\approx\!-2.5$, one notices an abrupt increase in $n_{\text{B}}$ accompanied with a breakdown of St\v{r}eda's formula [Fig.~\ref{fig_FCI}(c)], as well as a sudden drop in the many-body gap [Fig.~\ref{fig_FCI}(d)]:~all these signatures point towards a phase transition (gap-closing point) in the thermodynamic limit. Due to the finite size, the minimal many-body gap is found to be $\Delta=0.016$, which suggests a realistic ramping time $\tau\!\sim\!100$ for adiabatic state preparation, compatible with recent experiments~\cite{2022Leonard}.
% $\epsilon_R=-3$finite size gap in superfluid?

We now analyze how an FCI ground state can be dynamically prepared by slowly ramping up the reservoir energy to the ideal value $\epsilon_R\!\approx\!-2$. To optimize adiabatic preparation, we adjust the ramp according to the many-body gap; see the inset in Fig.~\ref{fig_FCI}(d). By tracking the bulk density during the ramp [Fig.~\ref{fig_FCI}(e)], one recovers the formation of a plateau for a sufficiently long ramping time $\tau$, in agreement with the adiabatic-limit prediction of Fig.~\ref{fig_FCI}(c). As further confirmed by the local St\v{r}eda marker, an FCI ground state with $C_{\text{str}}\!\approx\!0.5$ is prepared for adiabatic times $\tau\gtrsim 140$ [Fig.~\ref{fig_FCI}(f)]. The comparison with a less efficient, linear ramp is presented in Ref.~\cite{supp}.

\paragraph{State preparation via repeated cleaning.}
So far, we have discussed a protocol by which particles are injected from a reservoir into an empty system. Motivated by the ability of easily preparing an empty reservoir (a trivial zero-entropy state), we now explore the possibility of using the reservoir as a vacuum-cleaning resource in view of preparing ground states in the system. As sketched in Fig.~\ref{fig_sketch}(c), one considers a `dirty' (excited) initial state within our system. The cleaning cycle is then as follows:~(i) we slowly lower the reservoir energy, such as to retrieve excitations (hot atoms) from the system in a controlled manner; (ii) after this cleaning process, one rapidly lifts the reservoir until it becomes decoupled from the system; and (iii) one completely empties the reservoir. This cleaning cycle is then repeated $n_{\text{cyc}}$ times, until convergence is reached towards a target insulating (QH) state within the box. The advantage of this scheme is two-fold:~the empty reservoir state can be viewed as a perfect and easy-to-prepare zero-temperature state of holes; the difficulty in removing particles that are located deep in the bulk is compensated by several repetitions. 
%The aim is to remove the particles occupying higher energy states so that the left particles in the system occupy the lowest band characterized with Chern number $C=1$.
%another scheme about removing excitation from the system for the purpose of state preparation,

%Since the system is now almost completely decoupled from the reservoir, one can easily remove the particles in the reservoir. The ability of preparing such an empty reservoir makes it possible to have several cycles.

\begin{figure}
	\centering\includegraphics[width=0.99\linewidth]{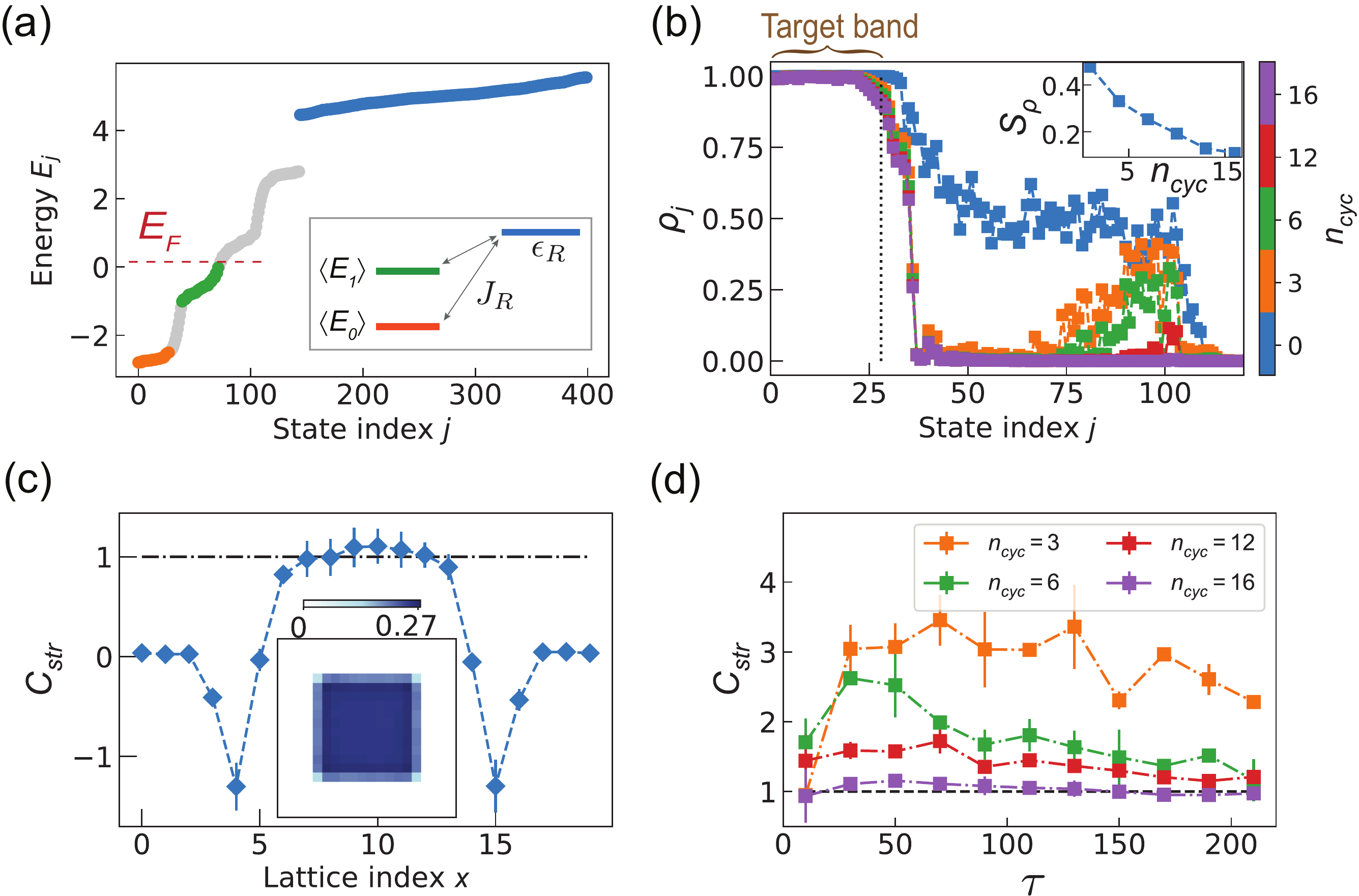}
	\caption{Preparing a Chern insulator by using the lift-removal strategy. (a) Energy spectrum of the HH model coupled to a trivial reservoir. Here, we partition a lattice of size $20\times20$ into a target $12\times 12$ system (central box) and a surrounding reservoir. $E_F$ denotes the Fermi energy. We set a flux  $\phi\!=\!\pi/2$ in the system only. Inset:~Simplified three-level model; $\left\langle E_{0}\right\rangle$ and $\left\langle E_{1}\right\rangle$ denote representative energies of the lowest two bands. (b) Population in HH eigenstates $\rho_j$ for different $n_{\text{cyc}}$. Inset: the HH orbital entropy versus $n_{\text{cyc}}$. We set $J_R\!=\!0.15$ and $\tau\!=\!90$ per cycle. (c) The local St\v{r}eda marker of the evolved state at $n_{\text{cyc}}\!=\!16$, as a function of the site index along the middle row. Inset: the corresponding spatial density distribution of the evolved state. (d) The local marker as a function of $\tau$ for different cycles $n_{\text{cyc}}$. The local marker is averaged over a disk at the center with a radius of $r\!=\!2$. The error bars denote the standard error of the regression slope used to extract $C_{\text{str}}$.}  % $\left\langle E_{0}\right\rangle \simeq-2.7$, $\left\langle E_{1}\right\rangle \simeq-0.6$ 
	\label{fig_prep}
\end{figure}

We apply this scheme to a concrete preparation sequence, designed to prepare Chern insulators in atomic HH systems. Inspired by Ref.~\cite{2015Aidelsburger}, we start from a trivial metal realized by loading non-interacting fermions in a square lattice at half-filling in the presence of a staggered potential~\cite{supp}. We then ramp up the flux in the lattice to the value $\phi\!=\!\pi/2$, while reducing the staggered potential, hence changing the topological nature of the bands:~at the end of this sequence, the target lowest band has a Chern number $C\!=\!1$. Due to the occupation of higher bands in the initial metallic state, the target (lowest) band remains perfectly filled during the whole duration of this sequence, despite the gap closing ($C\!=\!0\!\rightarrow\!1$). The (irregular) band populations, obtained at the end of this sequence, are shown by blue dots in Fig.~\ref{fig_prep}(b).

%we partition a square lattice of size $20\times20$ into a target system with size $12\times 12$ within a center box and a reservoir being the rest. We set $J_{\ell\ell'}=J,\phi_{\ell\ell'}=\phi n,\epsilon_{\ell}=[(-1)^{m}+(-1)^{n}]\delta/2$ for the system, and $J_{\ell\ell'}=J_{R},\phi_{\ell\ell'}=0,\epsilon_{\ell}=\epsilon_{R}$ for the reservoir. Here, $(m,n)$ are the lattice indices of site $\ell$ along $x$ and $y$ direction respectively and $\delta$ is the strength of staggered potentials.
%To be more realistic, we use an easily-prepared metal as our initial state with $\phi_i=0,\delta_i=20$. By linearly ramping up flux $\phi$ from 0 to $\pi/2$ and decreasing $\delta$ to 0 in the system, one arrives at a state with many higher band occupations~\cite{2015Aidelsburger}, as shown by the blue dots in Fig.~\ref{fig_prep}(d). 
Our aim is to remove atoms from higher bands, while leaving the lowest Chern band ($C\!=\!1$) almost perfectly filled, in view of forming a CI in the system. To achieve this goal, we now apply our vacuum-cleaning protocol by dynamically tuning the reservoir energy $\epsilon_R$. During each cycle of duration $\tau$, we vary $\epsilon_R(t)$ with a saturation function~\cite{supp}, using a large initial value $\epsilon_R^i=4$. At the end of this first cycle, $\epsilon_R(t)$ reaches the value $\epsilon_{R}^f\!=\!-1.14$, which is located right below the first excited Bloch band of the system. After lowering the reservoir to the final value $\epsilon_R^f$, we then quickly lift it up until it becomes effectively decoupled from the system; we then empty the reservoir and complete one cycle. We then repeat this cleaning sequence, but for the sake of efficiency, we progressively increase the final value $\epsilon_{R}^f$ at each cycle to properly address all the higher bands~\cite{supp}. We note that this cleaning scheme can be understood through a simplified 3-level toy model [inset of Fig.~\ref{fig_prep}(a)], which can serve as a guide in view of optimizing the control parameters~\cite{supp}.

%modify the final value of $\epsilon_R^f$ at each cycle $n_{\text{cyc}}$ according to
%\begin{equation}
%\epsilon_{R}^f(n_{\text{cyc}})=\epsilon_{R}^f(1)+J_{R}\xi\frac{n_{\text{cyc}}-1}{N_{\text{cyc}}-1},
%\end{equation}
%with $n_{\text{cyc}}=1,\cdots,N_{\text{cyc}}$ and $N_{\text{cyc}}$ is the total cycle number. The coefficient $\xi$ controls the final value of $\epsilon_R$.  % as shown by the dotted-dashed line in Fig.\ref{fig_prep}(d). 

Figure~\ref{fig_prep}(b) demonstrates the efficient depletion of the excited bands (and the resulting decrease of entropy) as a function of the cycle number $n_{\text{cyc}}$. In this process, the bulk states of the lowest band remain almost perfectly filled, and we find that a satisfactory CI ground state is formed after 16 cycles of duration $\tau\!=\!90$. We plot the local (single-site) St\v{r}eda marker in Fig.~\ref{fig_prep}(c), which confirms the topological nature of the bulk, $C_{\text{str}}\!\approx\!1$. We plot $C_{\text{str}}$ as a function of the ramping time per cycle for different $n_{\text{cyc}}$ in Fig.~\ref{fig_prep}(d). This shows that an efficient cleaning is reached for $n_{\text{cyc}}\!\approx\!12$ cycles of duration $\tau\!\gtrsim\!200$, and for $n_{\text{cyc}}\!\approx\!16$ cycles of duration $\tau\!\gtrsim\!50$.

%The advantages are that (i) the empty (zero temperature) reservoir is easy to prepare; (ii) preparing an empty reservoir again and again make it possible to have several cycles.

%\section{Experimental feasibility??}

\paragraph{Concluding remarks.}
%We have shown that dynamically tuning box potentials offers a versatile tool for the purpose of preparing nontrivial (strongly-interacting) states. We found that injecting particles from the reservoir into the system can be used not only to prepare topologically nontrivial ground states, but also to populate chiral edge states in an energy-selective manner. Motivated by the ability of easy tuning of box potentials and controlled removal of particles, we devise a lift-removal protocol, akin to the evaporative cooling,  so as to prepare nontrivial states like Chern insulators. Remark that while we explore a system-reservoir configuration that involves a coupling through particle hopping across a boundary, other designs (such as using a two-layer geometry or two coupled internal states of atoms) can also be envisaged. Our work offers a new path towards realizing and studying topologically nontrivial states with optical box traps.

This work explored different possibilities offered by the design of tunable boxes in cold-atom experiments, setting the focus on the realization of topological states. This approach offers substantial advantages:~it relies on the preparation of a simple initial state in the reservoir and the ability to dynamically tune the latter's energy relative to the system region. In this sense, our open-system approach does not require any fine-tuning nor complicated time-dependence of the system parameters, and it is readily applicable to create a broad class of (bosonic or fermionic) many-body states of interest, including exotic Mott insulators and antiferromagnetic states in Hubbard-type models. While we considered a spatial separation between the system and reservoir regions on the 2D plane, we note that a double-layer configuration could also be envisaged to further enhance the transfer of particles between the two regions; this could be realized using a bilayer optical lattice or by exploiting two laser-coupled internal states of an atom. Finally, it would be interesting to combine the injection and cleaning schemes presented in this work, in view of realizing large FCI states or to explore quantum thermodynamics.

\emph{Acknowledgments.}
The authors thank Julian Leonard, Yanfei Li, Nir Navon, Cecile~Repellin, Raphael~Saint-Jalm, Perrin ~Segura, Boye Sun, Amit~Vashisht and Christof Weitenberg for discussions. J.D. acknowledges the support of the Solvay Institutes, within the framework of the Jacques Solvay International Chairs in Physics. Work in Brussels is also supported by the FRS-FNRS (Belgium), the ERC Starting Grants TopoCold and LATIS, and the EOS project CHEQS. M.A. and A.E. acknowledge support from the Deutsche Forschungsgemeinschaft (DFG) via the Research Unit FOR 2414 under Project No. 277974659. M.A. also acknowledges funding from the DFG under Germany’s Excellence Strategy – EXC-2111 – 390814868.

%\bibliographystyle{apsrev4-1}
%\bibliography{mybib}


\clearpage
%%%%%%%%%%%%%%%%%%%%%%%%%%%%%% \appendix %%%%%%%%%%%%%%%%%%%%%%%%%%%%%%%%%%%%%%%%%%

\onecolumngrid
\setcounter{equation}{0}
\setcounter{figure}{0}
\renewcommand{\theequation}{S\arabic{equation}}
\renewcommand{\thefigure}{S\arabic{figure}}
\renewcommand{\thesection}{S\arabic{section}}
\renewcommand{\thesubsection}{\thesection.\arabic{subsection}}
\renewcommand{\thesubsubsection}{\thesubsection.\arabic{subsubsection}}

\begin{center}
	{\Large\bfseries Supplementary Material}
\end{center}

\section{I.~~ Brief reminder of the Harper-Hofstadter band structure: \\ Edge states and Chern numbers}
In this work, we illustrate the cold-atom elevator scheme using the paradigmatic Harper-Hofstadter (HH) model, which describes hopping of particles on a two-dimensional square lattice with a homogenous magnetic flux $\phi$ per plaquette. The HH Hamiltonian can be written in the Landau gauge as %~\cite{1955Harper,1976Hofstadter}
\begin{equation}
    \hat{H}=-J\sum_{m,n}\left(e^{-i\phi n}\hat{a}_{m+1,n}^{\dagger}\hat{a}_{m,n}+\hat{a}_{m,n+1}^{\dagger}\hat{a}_{m,n} \text{+h.c.}\right),
\end{equation}
where the operators $\hat{a}_{m,n}$ ($\hat{a}^\dagger_{m,n}$) annihilate (create) a particle on the lattice site $\ell=(m,n)$, where $m$ and $n$ denote the site indices along the $x$ and $y$ directions, respectively.
The uniform flux per plaquette $\phi=2\pi \alpha$ is defined modulo $2\pi$. For a rational flux $\alpha=p/q$, with $(p,q)$ prime numbers, the single-particle spectrum splits into $q$ subbands. When represented as a function of the flux $\alpha$, these energy bands form a fractal structure known as the Hofstadter butterfly. 

\begin{figure}[H]
	\centering\includegraphics[width=0.9\linewidth]{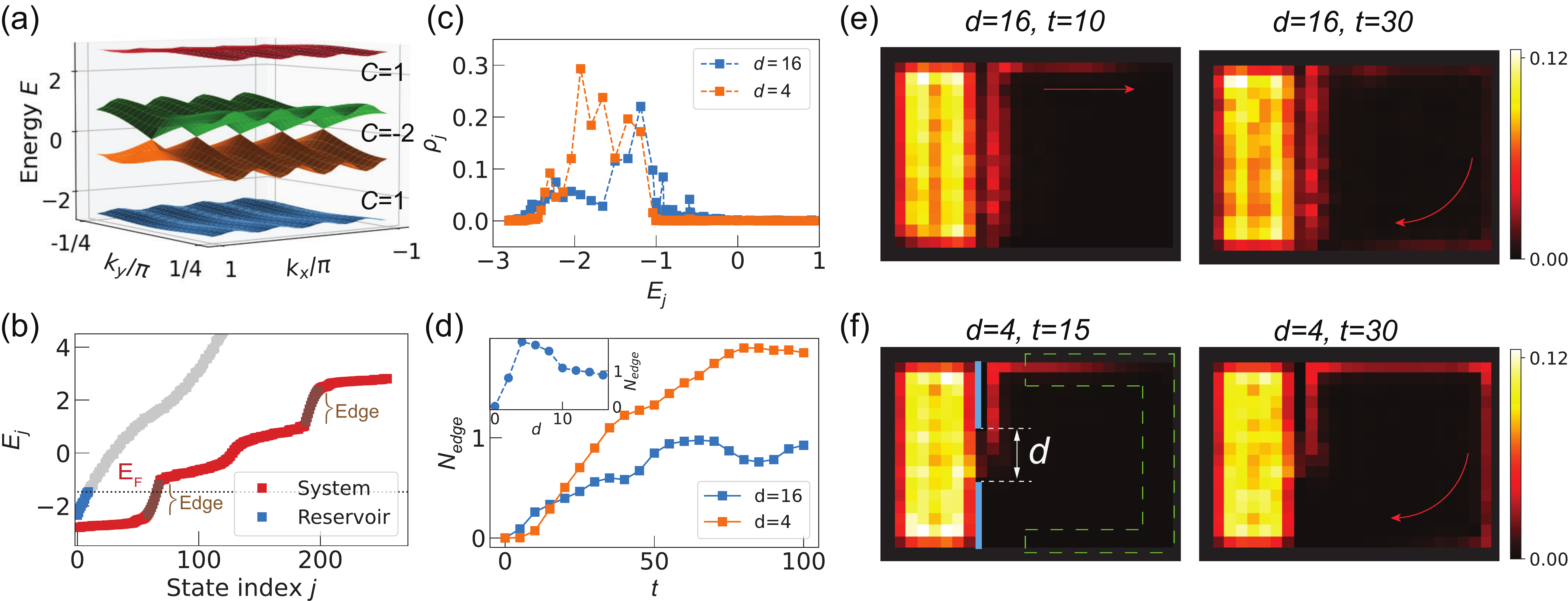} %supp_edge_png/eps
	\caption{ \textbf{Edge state injection with a trivial reservoir.} (a) Energy spectrum as a function of quasimomenta $k_x$ and $k_y$. We set the lattice constant $a=1$. The Chern numbers of the separated bands are $C=1, -2, 1$, respectively. (b) Energy spectrum as a function of the state indices. The blue and red dots correspond to the Hamiltonian describing the trivial reservoir (with chemical potential $\epsilon_R=1.5$) and the Hofstadter system, respectively. We consider a trivial reservoir of size $7\times16$ which is coupled to a system of size $16\times 16$ with flux $\phi=\pi/2$ per plaquette. A high wall (one site wide) is placed between the reservoir and the system [blue lines in (f)], which leaves a door of width $d$ open in the middle. (c) Population of HH eigenstates $\rho_j$ as a function of their energy $E_j$ for different door width $d$, at time $t=40$. (d) Number of the injected particles in the edge as a function of the time for different door width $d$. Inset: The number of particles in the edge as a function of $d$ at $t=100$. Snapshots of spatial density distribution for (e) $d=16$, (f) $d=4$. The green dashed line indicates the region used to count the number of particles at the edge, and the blue bonds represent the high wall. Initially there are $N=10$ particles in the reservoir, and the on-site potential in the reservoir is suddenly lifted to the value $\epsilon_R\!=\!1.5$. Energy and time are in units of $J$ and $\hbar/J$, respectively. The arrows in (e) and (f) are a guide to the eye for the chiral motion.} 
	\label{supp_edge}
\end{figure}

The finite (uniform) flux per plaquette $\phi$ breaks time-reversal symmetry, hence leading to non-trivial topological properties:~the Hofstadter bands are associated with a non-zero (first) Chern number. Considering the flux $\alpha=1/4$ as an example, the HH model exhibits 4 subbands, with the middle two bands touching at singular (Dirac) points [Fig.~\ref{supp_edge}(a)]. The lowest and the highest bands share the same Chern number $C=1$; while the middle degenerate band is associated with the Chern number $C=-2$. 

Under open boundary conditions, the model hosts chiral edge modes whose energies are located within the bulk energy gaps; these edge states can be identified in the red spectrum displayed in  Fig.~\ref{supp_edge}(b), where a steeper slope reflects the low density of (edge) states within the bulk gaps. Considering the flux $\alpha=1/4$ (main text), the two bulk gaps each hosts a single edge mode. The chirality of these two edge modes are opposite:~the edge mode located in the lowest (resp. highest) bulk gap propagates in a clockwise (resp. anti-clockwise) manner; see also Fig. 2(d) and (f) in the main text, where these two opposite chiral motions are revealed using the elevator scheme.

\section{II.~~ Edge-state injection using a trivial reservoir}

In the main text, we have shown that chiral edge states can be populated in an energy-selective manner, using a system-reservoir setting immersed in a uniform magnetic flux. In this Appendix, we show that a trivial reservoir (i.e.~a reservoir without flux) can also be used to inject particles into the edge states of a quantum Hall system. 

Similarly to the sketch in Fig.~\ref{fig_sketch}(a) in the main text, we couple a system described by the HH model to a reservoir without magnetic flux. Initially, the reservoir is filled with free fermions, and the system is empty. Suddenly lifting the reservoir on-site potential $\epsilon_R$, such that the energy of the particles in the reservoir becomes resonant with an edge mode of the system [Fig.~\ref{supp_edge}(b)], is found to generate chiral edge currents in the system [Fig.~\ref{supp_edge}(e)]. Even though only a small fraction of particles are injected into the edge states in this case, we find that applying a high wall between the reservoir and the system, but leaving a small door open in the middle of the wall, can improve the efficiency. In Fig.~\ref{supp_edge}(f), we plot the snapshots of density distributions of the time-evolved state using a small door opening of width $d\!=\!4$, from which we observe stronger chiral edge currents appearing in the system. We attribute the better efficiency of the door-opening configuration to the fact that a wider range of quasi-momenta is offered by a smaller door width, thus allowing for the population of more edge states in the system; see  Fig.~\ref{supp_edge}(c) where we plot the population of HH eigenstates for different $d$.

By monitoring the number of particles $N_{\text{edge}}$ injected in a specific region of the edge [depicted by the dashed lines in Fig.~\ref{supp_edge}(f)], we find that tuning the door width $d$ manifests as a way of controlling the number of particles injected into the edge states, as shown in Fig.~\ref{supp_edge}(d). In the inset of (d), we plot $N_{\text{edge}}$ as a function of the door width $d$ at time $t=100$, and the optimal fraction of injected particles is found at a door width of intermediate size.

\section{III.~~ Chern insulator preparation using the injection scheme}
\begin{figure}
	\centering\includegraphics[width=0.9\linewidth]{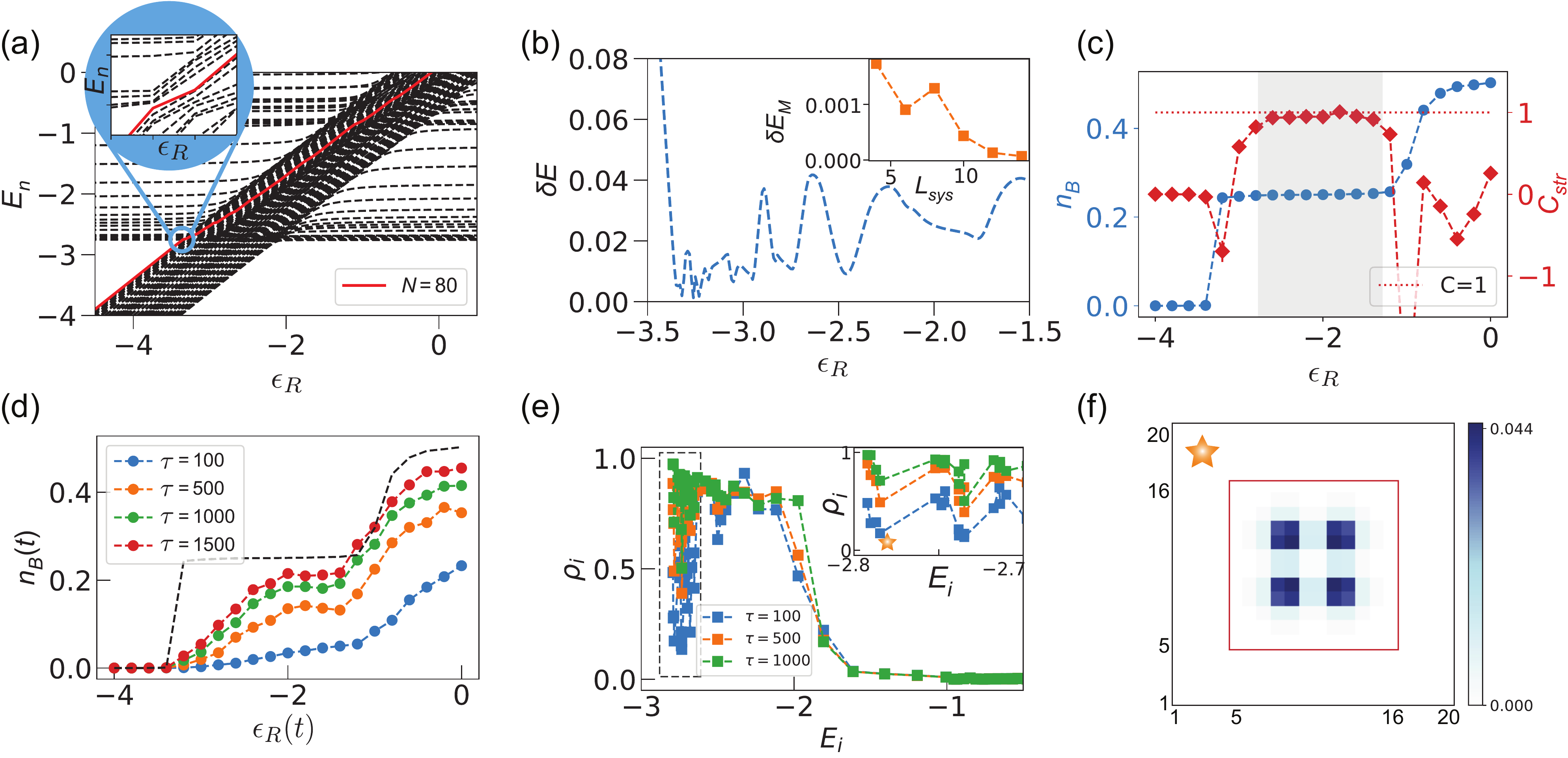}
	\caption{ \textbf{Preparing a Chern insulator based on injection.} (a) Energy spectra of the entire setup as a function of $\epsilon_R$. The red line indicates the Fermi level. Considering a box of size $12\times12$ with a target system being a $8\times8$ sub-box at the center, the Fermi level corresponds to the 80th eigenvalue. (b) Energy gap (difference between the 80th and 81st eigenvalues) for $J_R\!=\!0.2$. The minimal energy splitting is found to be $\delta E_M\!=\!0.0013$. Inset:~the minimal energy splitting $\delta E_M$ as a function of the system length $L_{\text{sys}}$; here, we consider square boxes of increasing size $L\times L$, and keep the ratio $L/L_{\text{sys}}\!\simeq\!1.6$ constant. (c) The bulk density and the local St\v{r}eda marker as a function of $\epsilon_R$. The bulk density is evaluated within the central $2\times 2$ sites. The shadow indicates the regime of Chern insulator ground states in the system. (d) The bulk density as a function of $\epsilon_R$ for different $\tau$. The black dashed line is the ground state density shown in (c), which represents the adiabatic limit. Here we linearly ramp up $\epsilon_R$ from $-4$ until $0$ within time $\tau$, with the initial state being a fully filled trivial reservoir. (e) The population of the eigenstates of the Hofstadter model. Here we consider a bigger box of size  $20\times20$ partitioned into the target system of size $12\times12$ and a reservoir. We ramp up $\epsilon_R$ from $-4$ until $-2.5$ by using a saturation function with time $\tau$. Inset: zoom-in of heavy depletion region indicated by the dashed box. (f) The spatial density distribution of a typical isolated bulk state with index $j=3$. Here, the tunneling strength is $J_R=0.2$.  } 
	\label{supp_inj_CI}
\end{figure}

Here, we investigate the possibility of preparing a CI based on injecting particles from a reservoir into an empty system.
As an example, we partition a large box of size $12\times12$ into a sub-box of size $8\times8$ at the center (the ``system") and we define the rest as a trivial reservoir (without magnetic flux). The middle sub-box is our target system, described by the HH model with flux $\phi=\pi/2$ per plaquette. 
We first analyze the ground-state properties of such a system-reservoir setup. Figure~\ref{supp_inj_CI}(c) shows the bulk density $n_{\text{B}}$ as a function of the reservoir potential $\epsilon_R$. The incompressible nature of the CI ground state manifests as a plateau in the bulk density. Its topological nature can be confirmed from the local St\v{r}eda's marker shown by the red dots in Fig.~\ref{supp_inj_CI}(c), which forms a plateau around 1 in the region with bulk density saturation, which is in agreement with the Chern number $C=1$ of the occupied lowest band.

Based on this ground-state property, one can envisage to adiabatically prepare a CI by slowly ramping up the reservoir potential $\epsilon_R$. In our setup, such an adiabatic limit will be determined by the energy gap $\delta E$ just above the Fermi level. A typical avoided-crossing is shown in the inset of Fig.~\ref{supp_inj_CI}(a), and the energy gap $\delta E$ as a function of $\epsilon_R$ is plotted in Fig.~\ref{supp_inj_CI}(b). We find that, in order to build up a CI by ramping up $\epsilon_R$, one has to pass through an energy gap as small as $\delta E=0.0013$, which complicates the preparation in practice. The bulk density is plotted in Fig.~\ref{supp_inj_CI}(d) for different ramping times $\tau$, upon using a linear ramping up of $\epsilon_R$. It shows that thousands of tunneling times ($\sim 1/\delta E$) would be required to approach the adiabatic limit for this system size; see also the scaling shown in the inset of Fig.~\ref{supp_inj_CI}(b).

As regards the preparation of a large CI through particle injection, another difficulty concerns the presence of isolated eigenstates deep in the bulk of the system. Taking a box of size $20\times20$ as an example, we partition it into a target HH system of size $12\times12$ and define the rest as the reservoir. Considering an improved ramp (based on a saturation function), we obtain the eigenstate populations $\rho_j$ displayed in Fig.~\ref{supp_inj_CI}(e). From the zoom-in plot in the inset, we  identify several dips in the eigenstate populations:~those dips correspond to eigenstates that are located deep in the bulk, and which are essentially isolated from the reservoir. The spatial density distribution of such a typical isolated eigenstate is shown in Fig.~\ref{supp_inj_CI}(e), with the red box indicating the boundary between the target system and the reservoir. 

The existence of eigenstates located deep in the bulk, which are essentially decoupled from the reservoir, complicate the perfect filling of the target topological band. This issue can be solved by improving the system-reservoir coupling (e.g.~by increasing the system-reservoir interface or considering a double-layer configuration), or by considering systems with (strong) interactions; see main text and Section IV below.

\section{IV.~~ Fractional Chern insulator preparation using the injection scheme}

\subsection{A.~ Using a linear ramp}
Despite the difficulties mentioned above, injecting particles into an empty system can still be used to prepare a strongly-correlated FCI state, as demonstrated in Fig.~3 in the main text. The reasons are twofold. First, considering a small FCI droplet, the many-body gap across the injection process can potentially remain (reasonably) large due to finite size effects. Second, the strong interactions potentially prevent the isolation of bulk states, as previously illustrated in Figs.~\ref{supp_inj_CI}(e)-(f). As a further demonstration, here we show that a linear ramp of $\epsilon_R$ can also be used to prepare a small FCI droplet.

\begin{figure}[h]
	\centering\includegraphics[width=0.9\linewidth]{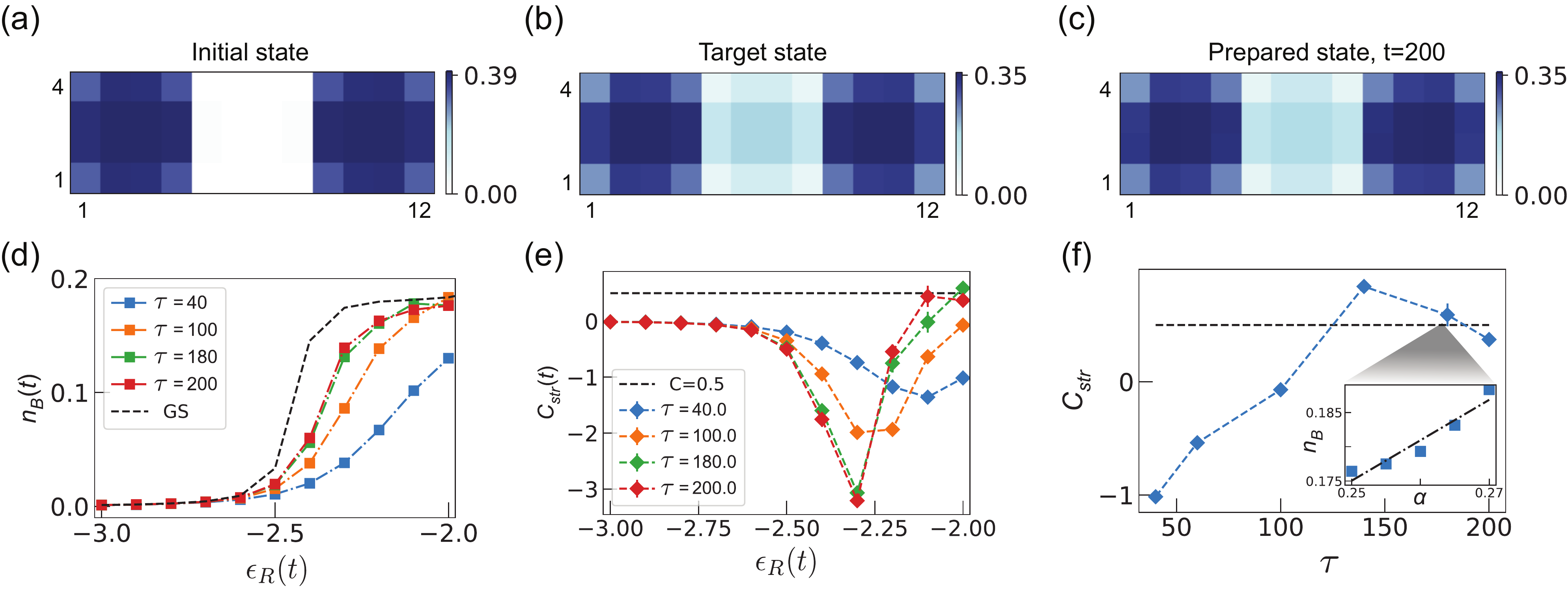}
	\caption{ \textbf{Preparing a fractional Chern insulator by using a linear ramp.} The spatial density distribution of (a) the initial state with $\epsilon_R=-3$, (b) the target state with $\epsilon_R=-2$ and (c) the prepared state at time $t=\tau=200$. (d) The bulk density as a function of $\epsilon_R$ for different $\tau$. The black dashed line corresponds to the adiabatic limit. (e) The local St\v{r}eda marker $C_{\text{str}}(t)$ as a function of $\epsilon_R(t)$ for different $\tau$. (f) The final St\v{r}eda marker as a function of $\tau$. Inset: linear fit of the density versus flux at $\tau=180$, which gives $C_{\text{str}}\approx0.59$. Here, we consider $N=12$ hard-core bosons, and the system of size $4\times4$ is coupled with two reservoirs with $J_R=0.15$, each of which has size $4\times4$.The error bars denote the standard error of a regression slope.} 
	\label{supp_FCI}
\end{figure}

We start from a trivial state with 6 hard-core bosons in each reservoir, as shown in Fig.~\ref{supp_FCI}(a), leaving the system empty. After linearly ramping up the reservoir potential $\epsilon_R(t)$ from $-3$ until $-2$ within a duration $\tau$, we track the bulk density, as evaluated within the central $2\times 2$ sites, as a function of $\epsilon_R(t)$; see Fig.~\ref{supp_FCI}(d). One observes that the bulk density approaches the adiabatic-limit prediction for a sufficiently long ramping time. The sudden increase of the bulk density indicates a transition. As further confirmed by the local St\v{r}eda marker in Fig.~\ref{supp_FCI}(e), the obvious dip (breakdown) of $C_{\text{str}}$ indicates a change in the topological properties. By using sufficiently long ramping times, $C_{\text{str}}$ gets close to the expected value $1/2$. The final St\v{r}eda marker as function of total ramping time is plotted in Fig.~\ref{supp_FCI}(f), with the inset showing that a ramping time $\tau=180$ leads to the marker's value $C_{\text{str}}\approx0.59$.

\subsection{B.~ Finite size effects}

After demonstrating the validity of our injection scheme for the preparation of a small FCI, we now analyze how the efficiency of this scheme scales with the size of the reservoir, as well as with the size of the target system.
Focusing on the sub-box configuration introduced in Fig.3(a) of the main text, we consider a target system of size $L_S\times L_S$ coupled with two identical reservoirs of size $L_R\times L_S$, where $L_S$ and $L_R$ denote the lengths of the system and reservoir, respectively. 

\begin{figure}
	\centering\includegraphics[width=0.99\linewidth]{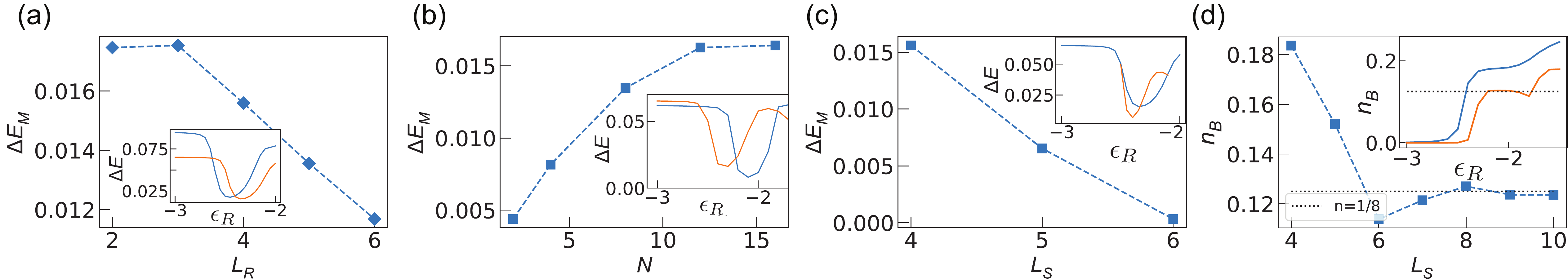}
	\caption{ \textbf{Finite size effects in the FCI-reservoir setup.} (a) The minimum gap $\Delta E_M$ as a function of the length of each reservoir $L_R$ for a fixed total number of particles $N=12$. Inset: The many body gap $\Delta E$ as a function of $\epsilon_R$ for $L_R$=3 (blue) and 4 (orange). (b) The minimum gap $\Delta E_M$ as a function of particle number $N$ for a fixed $L_R=4$. Inset: $\Delta E$ as a function of $\epsilon_R$ for $N$=4 (blue) and 12 (orange). In (a,b) we fix the system length $L_S=4$. (c) The minimum gap $\Delta E_M$ as a function of the system length $L_S$ for fixed $L_R=4, N=12$. Inset: $\Delta E$ as a function of $\epsilon_R$ for $L_S$=4 (blue) and 5 (orange). (d) The bulk density $n_B$ as a function of $L_S$ at $\epsilon_R=-2$ for $L_R=4, N=12$. Inset: $n_B$ as a function of $\epsilon_R$ for $L_S=4$ (blue) and $L_S=8$ (orange). The dotted line indicates the expected $n=1/8$ in the thermodynamic limit. The bulk density is averaged over a disk of radius $r=1$ at the center of the system. In (a-d), we choose the width of the whole setup to be $W=L_S$ and use $J_R=0.15, \alpha=1/4$. The dashed lines are the guide to the eye. } 
	\label{supp_FCI_size}
\end{figure}

We first investigate the effects of the reservoir length $L_R$ by fixing $L_S\!=\!4$ and the total particle number $N\!=\!12$. We calculate the many-body gap $\Delta E$ as a function of the reservoir energy $\epsilon_R$, within the range $-3\le\epsilon_R\le-2$, where a transition is known to occur (see Fig. 3 in the main text). From this, we extract the minimal many-body gap $\Delta E_M$, which sets the time-scale for adiabatic state preparation. The resulting quantity $\Delta E_M$ is plotted as a function of $L_R$ in Fig.~\ref{supp_FCI_size}(a). We find that for $L_R\ge3$, the minimal gap $\Delta E_M$ linearly decreases with $L_R$. Besides, by fixing $L_S\!=\!L_R\!=\!4$, we find a non-linear increase of $\Delta E_M$ as a function of the total particle number $N$; see Fig.~\ref{supp_FCI_size}(b). Since the value of the minimal gap $\Delta E_M$ determines the time scale for a possible adiabatic state preparation, the results presented here indicate that it is favorable to consider a reservoir of small size with many particles.

We now analyze how the size of the system $L_S$ influences the efficiency of the scheme, by fixing the reservoir length $L_R\!=\!4$ and the particle number $N\!=\!12$. As shown in Fig.~\ref{supp_FCI_size}(c), the minimal gap decreases with $L_S$, which signals the potential difficulty of preparing an FCI in a substantially larger lattice system. This limitation could be circumvent by optimizing the reservoirs (see above), or possibly, by combining the injection scheme with the cleaning method described in the main text. Another route for improvement could be offered by a double layered system-reservoir configuration; in practice, this could also be realized using two different internal states of an atom for the system and reservoir, respectively.

Last but not least, we analyze the (ideal) particle density expected within the bulk of the system $n_B$ as a function of the system size $L_S$, for a fixed reservoir energy $\epsilon_R\!=\!-2$; see Fig.~\ref{supp_FCI_size}(d). It is interesting to note that the particle density $n_B$ already approaches its thermodynamic-limit value $n_B=\nu\alpha=1/8$ (expected for a $\nu\!=\!1/2$ Laughlin state) for a system size $L_S\gtrsim8$. Specifically, we find $n_B\simeq0.127$ and  $n_B\simeq0.124$ for $L_S\!=\!8$ and $L_S\!=\!10$, respectively.

\section{V. State preparation via repeated cleaning}

\subsection{A.~ Rabi cycles}

The aim of this Appendix is to estimate optimal parameters for the vacuum-cleaning scheme, by analyzing a simplified few-level model.

In Fig.~4(a) of the main text, we presented a sketch of a simplified 3-level toy model for our system-reservoir setup. Our aim is to gain more intuition on our repeated cleaning protocol by analyzing the Rabi oscillations that occur between these levels upon coupling them. 
To be specific, we introduce the Level-0, Level-1, and Level-$R$ to denote our (target) lowest Chern band, the first excited band and the reservoir band, respectively. The corresponding characteristic energies are represented by $E_0$, $E_1$ and $\epsilon_R$ in Fig.~4(a). Level-0 and Level-1 have the same coupling strength $J_R$ with Level-$R$. % transition between Level-0 and 1 

Starting from the situation where both the lowest and the excited bands are fully occupied, we aim to activate a coupling that keeps the lowest band occupation essentially unaffected, while cleaning off the population in the excited bands as much as possible. Specifically, in terms of the 3-level toy model, we analyze the occupation in Level-0 when the transition from Level-1 to Level-$R$ is maximal. Here, we treat the 3-level toy model as two independent two-level systems:~one describing the coupling of the reservoir to Level-0 and the other describing the coupling of the reservoir to Level-1. 

In a two-level quantum system, it is well-known that the Rabi formula describes the transition from one level to the other. Starting from a state initially occupying the first level (denoted by $a$), the transition probability to the second level (denoted by $b$) reads
\begin{equation}
	P_{ab}(t)=\frac{\gamma^{2}}{\gamma^{2}+(E_{a}-E_{b})^{2}/4}\sin^{2}\left(\sqrt{\frac{(E_{a}-E_{b})^{2}}{4}+\gamma^{2}}t\right),
	\label{P_ab}
\end{equation}
where $E_a$, $E_b$ represent the level energies and $\gamma$ is the strength of coupling  between these two levels.

\begin{figure}
	\centering\includegraphics[width=0.95\linewidth]{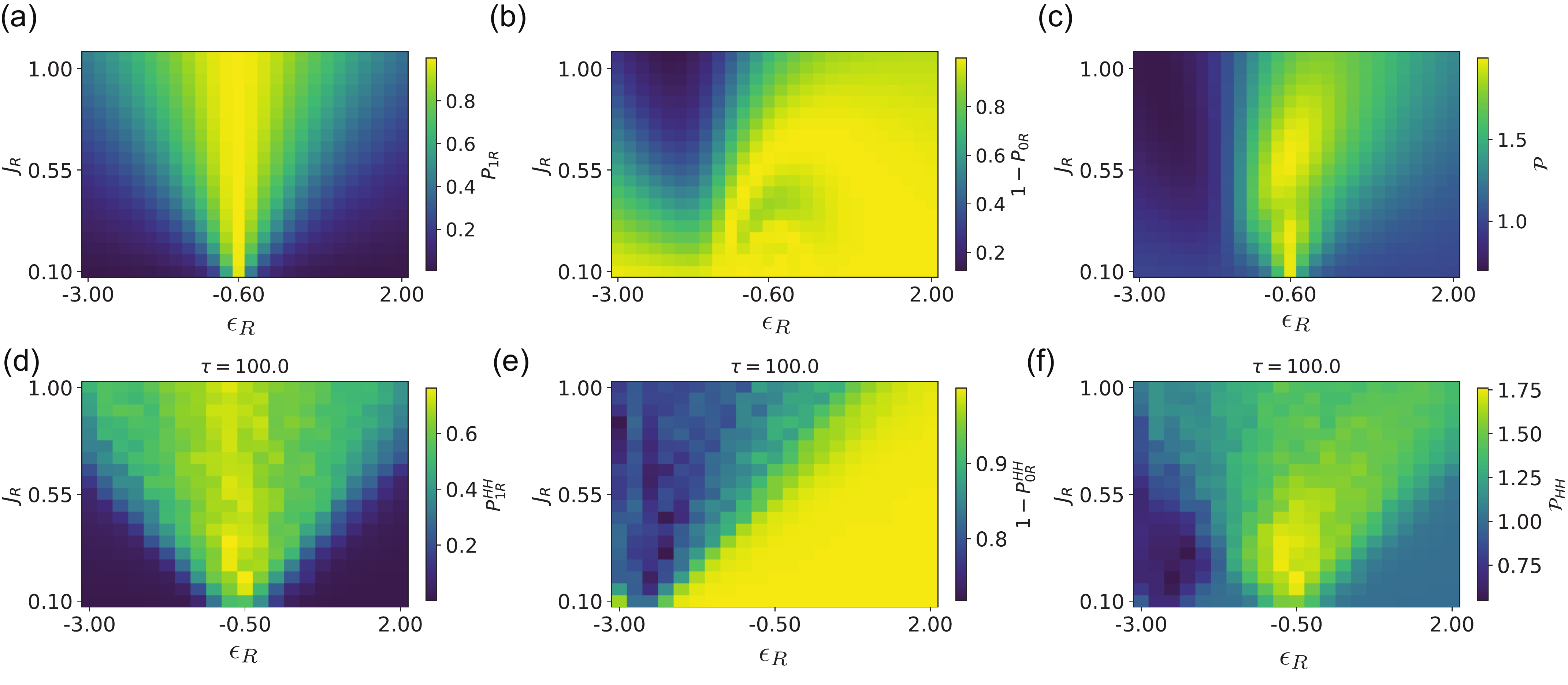}
	\caption{ \textbf{Rabi oscillations and quench dynamics.} (a) The maximal transition $P_{1R}^{\text{Max}}$, (b) the transition left in Level-0 $1-P_{0R}$ at $t=t^*$, and the characteristic population $\mathcal{P}$ of the simplified 3-level toy model. (d) The excited band transition $P_{1R}^{HH}$, (e) the remaining lowest band population $1-P_{0R}^{HH}$, and the characteristic population $\mathcal{P}_{HH}$ of the quench dynamics of our system-reservoir configuration at $t=100$. } 
	\label{supp_Rabi}
\end{figure}

We first analyze the transition from Level-1 to Level-$R$, and we establish its optimal regime by finding the maximal value $P_{1R}^{\text{Max}}=P_{1R}(t^*)$ using Eq.~\ref{P_ab}; the time $t^*$ will denote the optimal duration of the coupling. We plot $P_{1R}^{\text{Max}}$ for different values of $\epsilon_R, J_R$ in Fig~\ref{supp_Rabi}(a). When $\epsilon_R$ becomes resonant with $E_1\simeq-0.6$, which is taken as the mean value of the excited Hofstadter band, the transition probability $P_{1R}$ can take a value as large as 1. In order to study the remaining population in the Level-0 when $P_{1R}$ is maximal, we plot $1-P_{0R}$ at time $t=t^*$ in Fig~\ref{supp_Rabi}(b).  

Now, to identify an optimal regime for our control parameters, we define a figure of merit $\mathcal{P}$ as 
\begin{equation}
	\mathcal{P}= 1-P_{0R}+P_{1R}.
\end{equation}
In the ideal case where the population in Level-1 is completely removed ($P_{1R}=1$) while Level-0 remains perfectly unaffected ($P_{0R}=0$), one has the maximal value $\mathcal{P}=2$; vice versa one has the minimal value $\mathcal{P}=0$. As shown in Fig.~\ref{supp_Rabi}(c), it is preferable to set $\epsilon_R$ close to resonance with the excited band $E_1$ and to use a small coupling strength $J_R$, in order to empty the Level-1 while protecting the Level-0.

To make a further step towards our cleaning protocol, we now check the above-mentioned optimal parameter regime for a quench dynamics of our system-reservoir setup. Considering an initial state with $1/2$-filling per site within the HH system (the central sub-box), we investigate the quench dynamics by suddenly setting $\epsilon_R$ to a given value. Similarly, we now define the figure of merit $\mathcal{P}_{HH}$ for our HH system as, 
\begin{equation}
	\mathcal{P}_{HH}=1-P_{0R}^{HH}+P_{1R}^{HH}, 
\end{equation}
where we can express the possibility of transition from band $i$ to the reservoir in terms of the eigenstate populations $\rho_j$ as 
\begin{equation}
	P_{iR}^{HH}=\sum_{j=1}^{N_{i}}(1-\rho_{j})/N_{i},
\end{equation}
with $i=0,1$ and $N_i$ being the number of bulk states belonging to band $i$. %$N_{B0}$ and $N_{B1}$ being the number of bulk states belonging to the lowest and the second band, respectively. 
By plotting the figure of merit $\mathcal{P}_{HH}$ for different values of $J_R, \epsilon_R$ in Fig.~\ref{supp_Rabi}(f), one identifies an optimal parameter regime, where $\epsilon_R$ is close to the mean energy value $E_1$ of the excited band and where the coupling $J_R$ is relatively weak. In contrast, when $\epsilon_R$ is set close to the mean energy of the lowest band $E_0\simeq-2.7$, a small $\mathcal{P}_{HH}$ is found due to a heavy depletion of the lowest band. When $\epsilon_R$ is set to a very large value (off resonance), no transition takes place between the system and the reservoir, and thus $\mathcal{P}_{HH}$ saturates to 1. This analysis shows a qualitative agreement with the simplified Rabi model [Fig.~\ref{supp_Rabi}(c)]. 

\subsection{B.~ Dynamical cleaning scheme and scaling}

Despite the identification of an optimal parameter regime (see above), a simple quench dynamics does not lead to a satisfactory preparation of the CI, i.e.\ a complete depletion of the excited bands while leaving the lowest band perfectly filled.  This is essentially due to the finite dispersion of the HH bands. To solve this issue, we consider tuning $\epsilon_R$ in a time-dependent fashion, as we now explain. 

\begin{figure}
	\centering\includegraphics[width=0.9\linewidth]{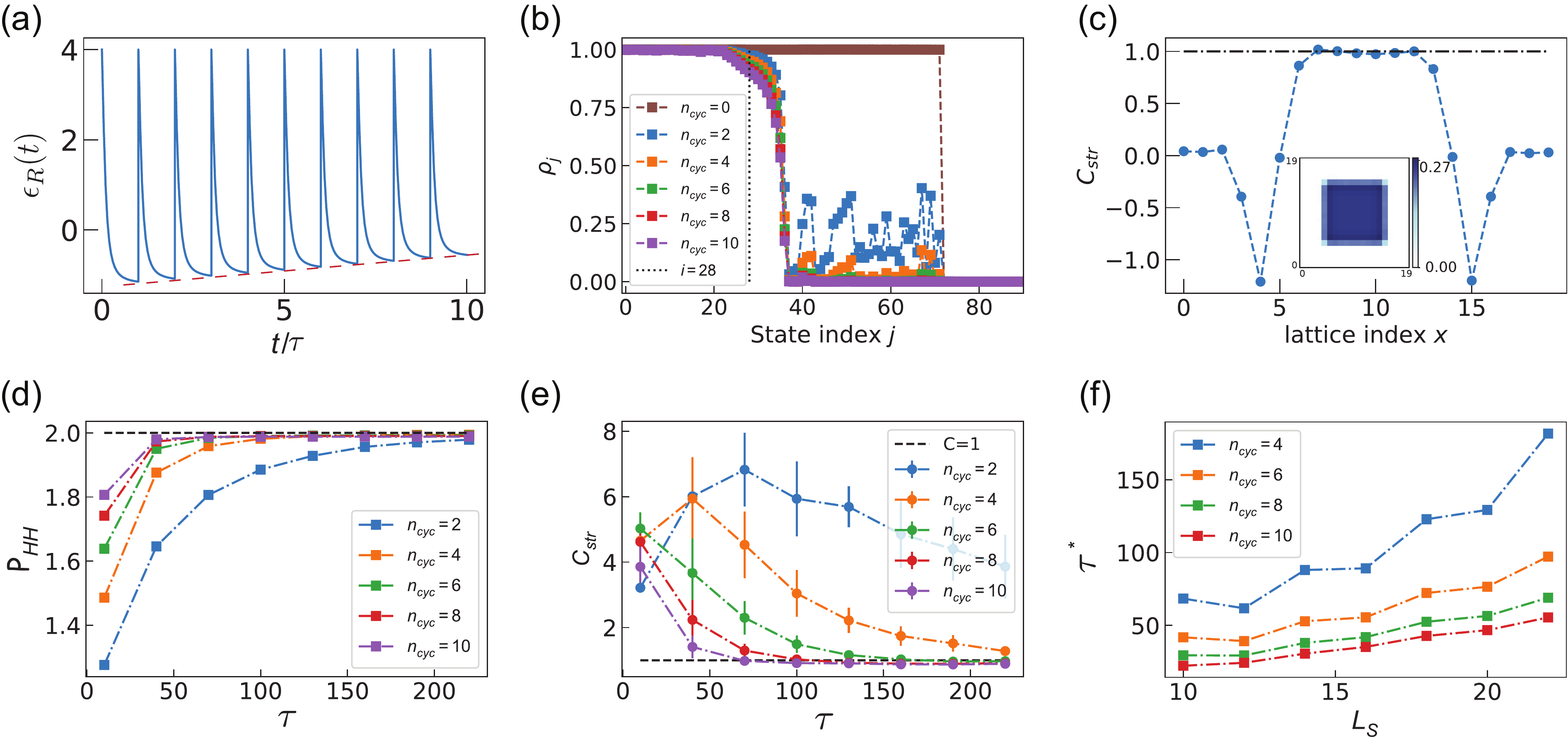}
	\caption{ \textbf{Preparing a Chern insulator based on cleaning.} (a) The repeated ramping protocol for $\epsilon_R$ according to a saturation function. The red dashed line is the guide to the eye. (b) Population in HH eigenstates for different $n_{\text{cyc}}$ with $\tau=70$ per cycle. (c) The local St\v{r}eda marker of the evolved state with $n_{\text{cyc}}=10, \tau=70$ as a function of lattice indices along the middle row. Inset: the corresponding spatial density distribution of the evolved state.  (d) The characteristic population $\mathcal{P}_{HH}$ as a function of $\tau$ for different cycles $n_{\text{cyc}}$. (e) The local marker as a function of $\tau$ for different cycles $n_{\text{cyc}}$ and $L_{S}$. The local marker is averaged over a disk at the center with a radius of $r=2$. The error bars denote the standard error of a regression slope. (f) The typical time $\tau^*$ as a function of the system length $L_S$. During the scaling, we consider square boxes and keep the length ratio between the whole set up and the system as $L/L_S\simeq1.6$.  For (b-e), we partition a lattice of size $20\times20$ into a target $12\times 12$ system (central box) and a surrounding reservoir with particle number $N=72$. We set a flux $\phi=\pi/2$ in the system and $\phi=0$ in the reservoir, and use $J_R=0.15, \xi=3$. } 
	\label{supp_CI_cleaning}
\end{figure}

We choose to dynamically tune the reservoir potential according to a saturation function,
\begin{equation}
	\epsilon_R(t)=\epsilon_{R}^i-(\epsilon_{R}^i-\epsilon_{R}^f)t/(\sqrt{c\tau^{2}+t^{2}}),
\end{equation}
with the initial value $\epsilon_R^i=4$, a constant parameter $c=0.02$ and the total ramping time $\tau$. As shown in Fig.~\ref{supp_CI_cleaning}(a), after reaching the final value $\epsilon_R^f$, we quickly lift $\epsilon_R$ to a high value. Since the system is now almost completely decoupled from the reservoir, one can easily remove the particles from the reservoir. The ability of preparing such an empty reservoir makes it possible to have several cycles. To increase the efficiency, one can design the final value of $\epsilon_R^f$ for each cycle $n_{\text{cyc}}$ as
\begin{equation}
	\epsilon_{R}^f(n_{\text{cyc}})=\epsilon_{R}^f(1)+J_{R}\xi\frac{n_{\text{cyc}}-1}{N_{\text{cyc}}-1},
	\label{mu_f}
\end{equation}
with $n_{\text{cyc}}=1,\cdots,N_{\text{cyc}}$ and $N_{\text{cyc}}$ is the total cycle number. The coefficient $\xi$ controls the final value of $\epsilon_R$. For the very first cycle, we take $\epsilon_{R}^f(1)=-1.14$ which is located right below the excited band.% as shown by the dotted-dashed line in Fig.\ref{fig_prep}(d).

Considering a many-body state at $1/2$-filling as our initial state, we now apply our repeated cleaning protocol to prepare a CI in the system:~we aim at removing the higher band populations but leaving the lowest Chern band almost perfectly filled. As shown in Fig.~\ref{supp_CI_cleaning}(b), we find an efficient depletion of the excited bands with the increase of the cycle number. During the ramp, the bulk states of the lowest band remain almost perfectly filled. This already points towards the realization of a CI state in the system. To further confirm its topological nature, we plot the local St\v{r}eda marker (averaged over the surrounding 4 sites) for $n_{\text{cyc}}=10$ with $\tau=70$ per cycle in Fig.~\ref{supp_CI_cleaning}(c), which shows a uniform bulk with $C_{\text{str}}\approx 1$. By plotting the figure of merit $\mathcal{P}_{HH}$, and the St\v{r}eda marker in Fig.~\ref{supp_CI_cleaning}(d) and (e), respectively, we find an efficient cleaning for $\tau\gtrsim70$ and $n_{\text{cyc}}\approx10$ cycles, or $\tau\gtrsim100$ and $n_{\text{cyc}}\approx8$ cycles. This illustrates how using more cycles would help reducing the total preparation time. Besides, in order to appreciate the scaling behavior, we plot a typical time $\tau^*$ corresponding to a figure of merit $\mathcal{P}_{HH}$ above 1.95 in Fig.~\ref{supp_CI_cleaning}(f). We find that the typical time per cycle increases more slowly with the system size as one increases the number of cycles. 

\subsection{C.~ Full preparation protocol using the cleaning scheme}

In the main text, we apply our repeated cleaning protocol to a realistic state, which has been prepared starting from a trivial metal (i.e. starting from a band structure with zero Chern numbers). Here we give more details about the full ramping protocol. 

\begin{figure}
	\centering\includegraphics[width=0.9\linewidth]{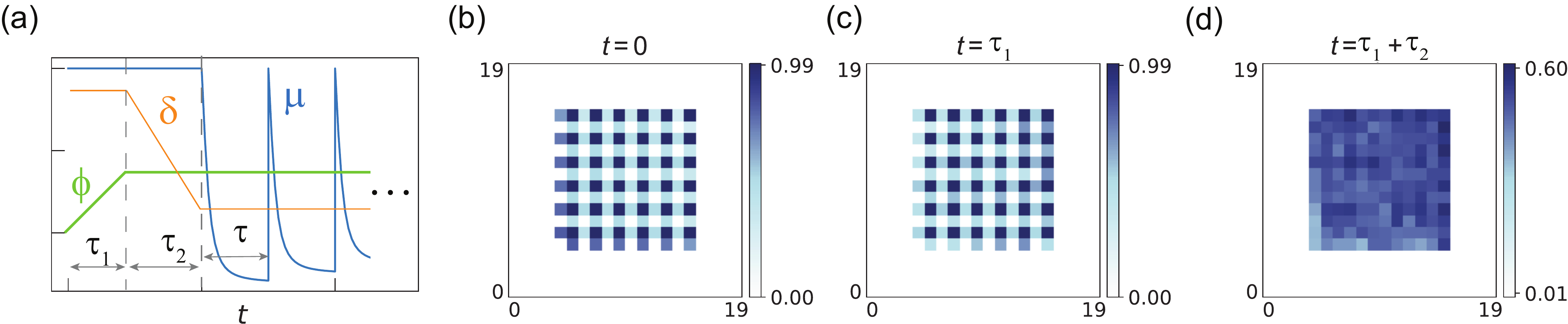}
	\caption{ \textbf{Full protocol for preparing CI based on cleaning} (a) The full protocol for ramping system parameters. The flux is linearly increased from 0 to $\pi/2$ within $\tau_1$, and then we decrease the staggered potentials $\delta$ from 20 to 0 within $\tau_2$. After that, we apply our cleaning protocol cycle by cycle with $\tau$ being the ramping time per cycle. (b-d) The spatial density distributions at $t=0, \tau_1$ and $\tau_1+\tau_2$, respectively. Here, we use time $\tau_1=\tau_2=100$ for tuning $\phi$ and $\delta$. } 
	\label{supp_full}
\end{figure}

We start from a trivial metal in the presence of a large staggered potential, $\delta_{\ell}=[(-1)^{m}+(-1)^{n}]\delta/2$, with $(m,n)$ being the lattice indices along $x$ and $y$ direction, respectively. We then linearly ramp up the flux in the lattice from 0 to the value of $\pi/2$. Since the system is still within a trivial regime (guaranteed by the large staggered potentials $\delta=20$), one can ramp up the flux rather quickly. After that, we slowly reduce the staggered potential in order to change the topological nature of the bands; this transition generates excitations in the higher bands for a realistic quasi-adiabatic ramp; see the sequence in Fig.~\ref{supp_full}(a). The density distribution of the evolved state, across the topological phase transition, is shown in Fig.~\ref{supp_full}(b)-(d); this latter situation corresponds to the blue dots in the band populations shown in Fig.~4(b) in the main text.

With this realistic (but highly excited) state in hand, we now apply our repeated cleaning protocol. In order to progressively address the higher bands at each cycle, we set $\epsilon_{R}^f(1)=-1.14$ and $\xi=14$ for the final value of $\epsilon_R^f$ defined in Eq.~(\ref{mu_f}); see also Fig.~\ref{supp_full}(a). In this case, the final value of $\epsilon_R^f$ in the last cycle reads 0.94, which is located near the top of the excited band. Note that even though we only consider a linear increase of $\epsilon_R^f$ at each cycle (for simplicity), a more efficient protocol could be designed by fine tuning $\epsilon_R^f$ within each cycle, e.g. by means of optimal control or even machine learning.

\end{document}